\documentclass[aip,graphicx]{revtex4-1}

\usepackage{graphics}

\draft 

\begin{document}

\title{Understanding the length dependence of molecular junction thermopower} 

\author{Olov Karlstr{\"o}m}
\author{Mikkel Strange}
\author{Gemma C. Solomon}
\email{gsolomon@nano.ku.dk}

\affiliation{Nano-Science Center and Department of Chemistry, University of Copenhagen, Universitetsparken 5, 2100 Copenhagen \O, Denmark}

\date{\today}

\begin{abstract}
Thermopower of molecular junctions is sensitive to details in the junction and may increase, decrease, or saturate with increasing chain length, depending on the system. Using McConnell's theory for exponentially suppressed transport together with a simple and easily interpretable tight binding model, we show how these different behaviors depend on the molecular backbone and its binding to the contacts. We distinguish between resonances from binding groups or undercoordinated electrode atoms, and those from the periodic backbone. It is demonstrated that while the former gives a length-independent contribution to the thermopower, possibly changing its sign, the latter determines its length dependence.  This means that the question of which orbitals from the periodic chain that dominate the transport should not be inferred from the sign of the thermopower but from its length dependence. We find that the same molecular backbone can, in principle, show four qualitatively different thermopower trends depending on the binding group: It can be positive or negative for short chains, and it can either increase or decrease with length.
\end{abstract}

%\pacs{73.63.-b, 73.50.Lw, 73.61.Ph, 82.35.Lr}

\maketitle

\section{Introduction}
In his seminal work, \textit{Physical Organic Chemistry},\cite{HammettBOOK} Louis P. Hammett writes: \textit{``It is one of the fundamental and most familiar assumptions of the science of organic chemistry that like substances react similarly and that similar changes in structure produce similar changes in reactivity. Yet the application of the principle requires so great an exercise of judgment, offers so wide an opportunity for the wisdom that comes only with experience and for the genius that seems almost intuition that there is some justice in the compliment or gibe, whichever it be, that this is an art not a science.''} This mindset carries over to nano science today and becomes the more general assumption that similar changes in structure produce similar changes in a physical observable. This assumption certainly holds true for the length dependence of single molecule conductance, an area undoubtedly outside the scope that Hammett envisaged in 1940. In this article, we show that we must not forget the caution implicit in Hammett's words in naively applying this principle to new physical observables, in this case the thermopower of molecular junctions. 

The thermopower (or Seebeck coefficient) is defined as, $S=-\Delta V/ \Delta T$, where $\Delta V$ is the bias required to prevent current from flowing through a system upon application of a temperature gradient $\Delta T$. The last five years have seen significant technological advances in measuring the thermopower 
of molecular junctions.\cite{Reddy2007,Malen2009,Tan2010,Widawsky2012,Baheti2008,Tan2011,Widawsky2013,Lee2013,Guo2013} Thermoelectric measurements indicate whether transport is dominated by the Highest Occupied Molecular Orbital (HOMO) or the Lowest Unoccupied Molecular Orbital (LUMO), information not easily accessible by measurements of the conductance, \cite{Paulsson2003} and are thus a useful complementary tool to understand charge transport.\cite{Baheti2008} Better understanding of thermoelectric transport will facilitate the interpretation of measurements, as well as enable the development of efficient thermoelectrics.\cite{Bergfield2009,Bergfield2010,Karlstrom2011} 

\begin{figure}[ht]
{\resizebox{!}{50mm}{\includegraphics{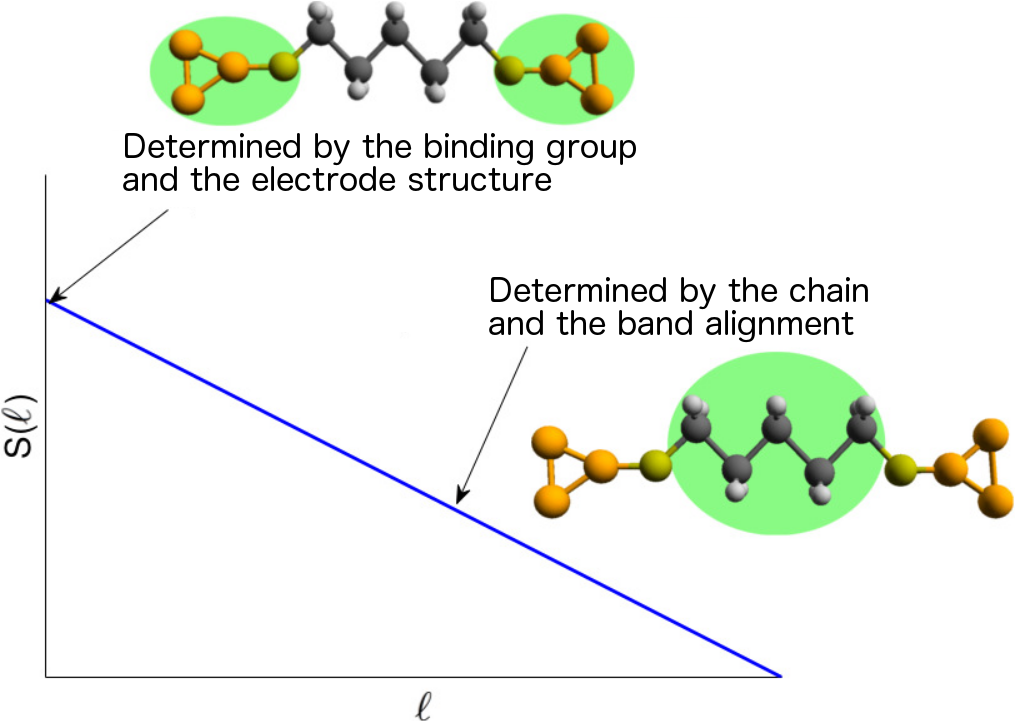}}}
\caption{The linear length dependence of thermopower. While the short chain value is set by the binding group and electrode structure, the slope is determined by the properties of the chain and the position of molecular backbone states relative to the Fermi level $E_F$.}
\label{pedagogic} 
\end{figure}

Previously, the length dependence of thermopower has been measured for alkane and phenyl chains,\cite{Malen2009,Tan2011,Widawsky2013} and theoretical investigations have been performed for both these systems.\cite{Tan2011,Pauly2008,Quek2011,Widawsky2013} It has been found that the conductance is simply exponentially suppressed with increasing molecular length,\cite{Li2008,Widawsky2013,Cui2002,Xu2003,Lee2005} independent of how the chain binds to the contact. As prior theoretical work has suggested that thermopower is less sensitive than conductance to details in the coupling to the electrodes,\cite{Paulsson2003} one would expect this to show less dependence on the binding group. However, the length dependence of the thermopower shows strong binding group dependence.\cite{Malen2009,Widawsky2013,Guo2013} Alkanedithiols have shown positive but decreasing thermopower,\cite{Malen2009} however,  an increasing trend has been observed in a recent publication.\cite{Guo2013} For alkanes directly coupled to gold a constant thermopower was reported.\cite{Widawsky2013} Thus, there are three qualitatively different results for three very similar systems. This paper aims at explaining how such different result can be obtained. Focus will be on the results of Malen et al.\cite{Malen2009} and Widawsky et al.,\cite{Widawsky2013} as these works report thermopower measurements for a larger set of chain lengths. The results of Guo et al.\cite{Guo2013} are discussed in the end of the paper. We use a minimal tight binding model and show that binding group induced resonances give a crucial length-independent thermopower contribution. Variation with length is governed by the position of molecular backbone states relative to the Fermi level. This is depicted in Fig.~\ref{pedagogic} which sketches the thermopower for alkanedithiols following the results of Malen et al.\cite{Malen2009} The figure indicates how different parts of the figure affect the thermopower trend.

\section{Results and Discussion}
\subsection{Theoretical Model}
\begin{figure*}[ht]
\begin{center}
{\resizebox{!}{35mm}{\includegraphics{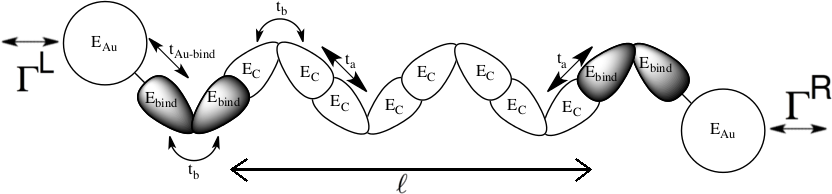}}}
\end{center}
\caption{Illustration of an alkane chain with length $\ell=4$ bonded to gold via binding groups. Each carbon atom and binding group atom have two orbitals, while the gold atoms only have one. The onsite energies and tunneling matrix elements are indicated in the figure, and the binding group orbitals have been shaded for clarity.  $\Gamma^L$ and $\Gamma^R$ denotes the couplings to the left and right contacts respectively.} 
\label{Fig2}
\end{figure*}

To investigate the transmission and thermopower of alkane chains, we used a simple Sandorfy-C-approximation model for the chain.\cite{Sandorfy1955} In this approximation, the hydrogen atoms are not included, i.e. only the $sp^3$-orbitals binding the carbons together are included. Each atom thus has two such orbitals. The coupling matrix element, from the $\sigma$-coupling between two orbitals on neighboring atoms, is given by $t_a$, while $t_b$ denotes the overlap between orbitals on the same atom, as shown in Fig.~\ref{Fig2}. The bandstructure for the periodic backbone of the chain can be derived by inserting a Bloch state ansatz in the Schr{\"o}dinger equation. This yields
\begin{equation}
E(k)=\pm\sqrt{t_a^2+t_b^2+2t_at_b\mathrm{cos}(k)},~0\leq k\leq \pi .
\label{bandstructure}
\end{equation}
The two bands will be referred to as the HOMO- and LUMO-band of the states extending over the backbone of the chain. We see that $t_a$ sets the splitting between the centers of the occupied and unoccupied bands, while the widths of the bands are, in the long chain limit, given by $2t_b$. As indicated in Fig.~\ref{Fig2}, the onsite energies of the carbon-, binding-, and gold-atoms are given by $E_C$, $E_{bind}$, and $E_{Au}$ respectively. For simplicity we assume that the coupling between the two orbitals on the binding group is equal to $t_b$, and that the coupling between the binding group and the carbon chain is $t_a$. The only difference between a binding atom and a carbon atom is thus the onsite energy. Finally, the coupling between the gold and the binding group is denoted $t_{Au-bind}$. The number of carbon atoms in the chain is labeled $\ell$, as indicated in Fig.~\ref{Fig2}. Note that this means that the number of atoms in the chain with binding groups is $\ell+2$. With these definitions we can write down a simple tight binding Hamiltonian $H$.

Assuming wide band contacts, the couplings to the contacts are characterized by the self-energies $\Sigma^{L/R}=-i\Gamma^{L/R}/2$, and the retarded Green's function can be calculated as
\begin{eqnarray}
G^r(E)=\left[ EI-H+\frac{i}{2}\left( \Gamma^L+\Gamma^R \right) \right]^{-1},
\label{Gr}
\end{eqnarray}
where the identity matrix $I$ is used as the overlap matrix, as we assume that we are working in an orthogonal basis. The transmission function is obtained as
\begin{eqnarray}
\tau(E)=\mathbf{Tr}\left[ \Gamma^LG^r(E)\Gamma^RG^a(E) \right],
\label{transmission}
\end{eqnarray}
where $G^a(E)$ is the advanced Green's function and is the hermitean conjugate of $G^r(E)$. Throughout the paper we will assume symmetric coupling to the electrodes, the magnitude of the coupling between the molecule and the left electrode, $\Gamma$, is equal to the coupling to the right electrode.

\subsection{Thermopower for exponentially suppressed transport}

\begin{figure}[ht]
\begin{center}
{\resizebox{!}{50mm}{\includegraphics{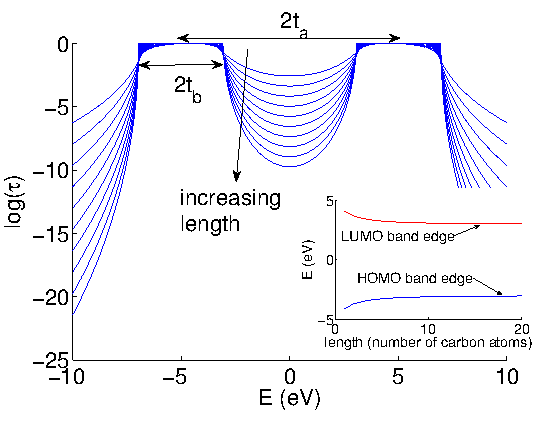}}}
\end{center}
\caption{The transmission functions for chains with $\ell=1...10$. Parameters are given by: $E_{bind}=E_{Au}=E_C=0$, $t_a=-5$~eV, $t_b=-2$~eV, $t_{Au-bind}=-5$~eV, and $\Gamma=4$~eV. The insert shows how the HOMO-LUMO gap closes with increasing length of the chain.}
\label{tau_tc0} 
\end{figure}

We will now look at what results can be derived from the model. We start by investigating how the different parameters of the model affect $\tau(E)$.  First, we study the dependence on the parameters for the carbon chain and for simplicity set $E_{bind}=E_{Au}=E_C=0$. The transmission functions for chains with $\ell=1...10$ and additional parameters given by $t_a=-5$~eV, $t_b=-2$~eV, $t_{Au-C}=-5$~eV, and $\Gamma=4$~eV are shown in Fig.~\ref{tau_tc0}. The temperature is assumed to be $T=300$~K throughout the paper.

As indicated in Fig.~\ref{tau_tc0}, $t_a$ sets the distance between the centers of the occupied and unoccupied bands, while $2t_b$ gives the width of the bands in the long chain limit. Throughout this paper we will refer to the highest occupied molecular orbital (HOMO) and lowest unoccupied molecular orbital (LUMO) as the highest and lowest eigenstates of the bands from the carbon backbone. In the case of the molecules bound directly to gold, these eigenstates are the HOMO and LUMO of a reduced system (terminal atoms removed). However, when we consider the system with binding groups, these orbitals are not necessarily the HOMO and LUMO of the full system and a reader should keep this in mind.

\begin{figure}[ht]
\begin{center}
{\resizebox{!}{130mm}{\includegraphics{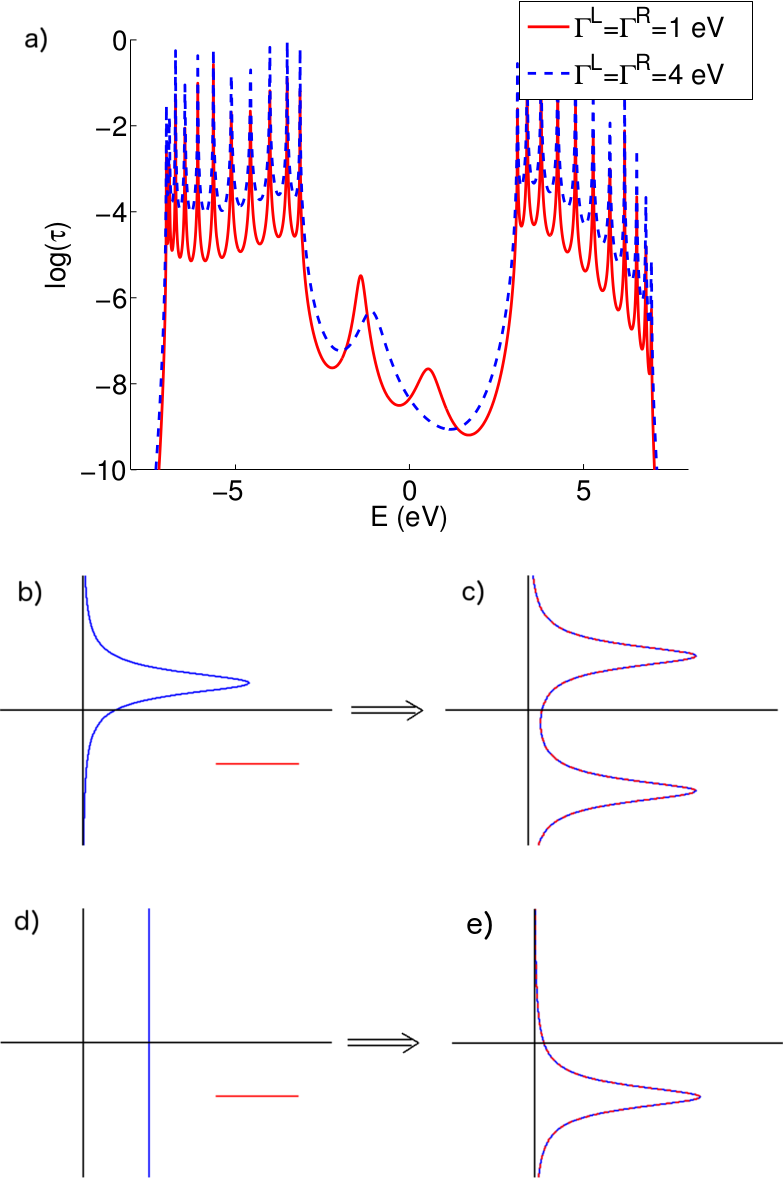}}}
\end{center}
\caption{a) $\tau(E)$ for the chains with $\ell=10$ and parameters $E_C=0$, $E_{bind}=-1$~eV, $E_{Au}=0$, $t_a=-5$~eV, $t_b=-2$~eV, and $t_{Au-bind}=-1$~eV. b) Sketch of DOS at the gold atom (blue) for small $\Gamma$, and a discrete eigenstate of the chain (red). c) Sketch for the DOS for the state resulting from hybridization between gold and the discrete eigenstate. Since the couplings to the contacts $\Gamma$ are small the peaks in c) are shifted. d) When $\Gamma$ is large the DOS at the gold atom is constant (blue). e) Hybridization between gold and the discrete eigenstate results in a DOS peaked at the energy of the discrete state.}
\label{bindinggroup} 
\end{figure}

In Fig.~\ref{tau_tc0}, a relatively strong coupling between the chain and the gold is assumed. As $t_{Au-bind}$ decreases the transmission in the HOMO-LUMO gap develops more structure as can be seen in Fig.~\ref{bindinggroup} a). Here $\tau(E)$ is shown for chains with $\ell=10$ and parameters the same as in Fig.~\ref{tau_tc0} except for $t_{Au-bind}=-1$~eV and $E_{bind}=-1$~eV. When $\Gamma=t_{Au-C}=1$~eV, two peaks are observed around $E=0$. As $\Gamma$ increases to $4$~eV only the lower peak remains. The reason for this is explained in the lower panel of Fig.~\ref{bindinggroup}. The chain with binding groups has two eigenstates in the HOMO-LUMO gap. These two states are symmetric and antisymmetric with respect to the middle of the chain and are commonly referred to as gateway states. These are mainly localized at the left and right binding group, and will hybridize with gold. When the coupling to the contacts ($\Gamma$) is reduced, as in the case of a gold adatom on a surface compared with an atom in a flat surface, the density of states (DOS) is peaked around the onsite energy of the gold, in this case chosen to be $E_{Au}=0$, as can be seen in Fig.~\ref{bindinggroup} b). The hybridization will result in the DOS shown in Fig.~\ref{bindinggroup} c), where the peaks are shifted compared with that seen in b) due to the peaked DOS of gold. When $\Gamma$ is larger, which is the situation when the binding group couples to a flat gold surface, the gold DOS is quite flat as in d). Hybridization between gold and the discrete state will then only broaden the discrete state, resulting in the DOS shown in e). This agrees with $\tau(E)\propto \rho^L(E)\rho^R(E)$, where $\rho^{L/R}$ are the DOS at the contact edges.\cite{Prodan2007} The result in Fig.~\ref{bindinggroup} a) can therefore be understood from the DOS shown in Fig.~\ref{bindinggroup} c) and e).  

Based on McConnell's  work,\cite{McConnell1961} the transmission through a single channel in the HOMO-LUMO gap can, to a good approximation, be written as
\begin{eqnarray}
\tau(E)=e^{-\beta(E)\ell}\alpha(E).
\label{tau}
\end{eqnarray}
Here $\beta(E)$, determined by the chain properties, dictates the exponential suppression, and $\alpha(E)$
is mainly determined by the binding group and electrode structure.

When the temperature $k_BT$ is much smaller than the energy scales at which $\tau(E)$ varies,
the Sommerfeld expansion gives the following expression for the thermopower S
\begin{eqnarray}
S=-\frac{\pi^2k_B^2T}{3e}\left. \frac{\tau'(E)}{\tau(E)} \right|_{E=E_F}.
\label{S}
\end{eqnarray}
\noindent Inserting Eq. (\ref{tau}) into Eq.~(\ref{S}) gives
\begin{eqnarray}
S=\frac{\pi^2k_B^2T}{3e}\left. \left( \beta'(E)\ell-\frac{\alpha'(E)}{\alpha(E)} \right) \right|_{E=E_F},
\label{S2}
\end{eqnarray}
\noindent which was previously derived by Pauly et al.\cite{Pauly2008} The length dependence of the thermopower is decided by $\beta'(E_F)$. Charge transport is maximally suppressed at $E_0$, defined as $\mathrm{arg}\{\mathrm{max}(\beta(E))\}$. For the model considered in this paper $E_0=E_C=0$. 

\begin{figure}[ht]
\begin{center}
{\resizebox{!}{45mm}{\includegraphics{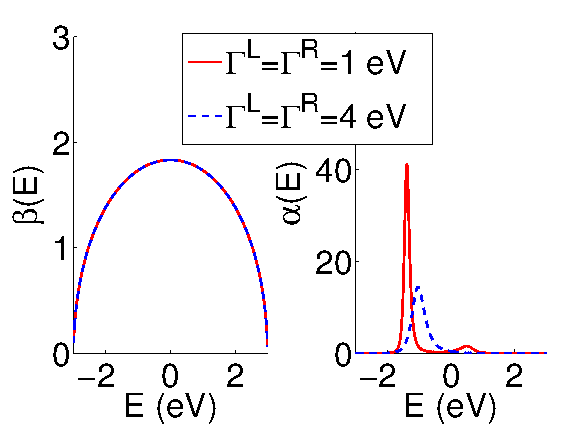}}}
\end{center}
\caption{$\beta(E)$ and $\alpha(E)$ for chains with $\ell=20$ directly coupled to gold. Parameters are the same as in Fig.~\ref{bindinggroup} a).}
\label{beta_alpha_pedagogic} 
\end{figure}

It is instructive to see the form of $\alpha(E)$ and $\beta(E)$ for the system corresponding to Fig.~\ref{bindinggroup} a). For short chains, $\ell\lesssim 4$, the transport is not exponentially suppressed along the chain and Eq.~(\ref{tau}) is therefore not appropriate in this limit. Thus, $\alpha(E)$ and $\beta(E)$ should be determined in the limit of long chains. In this paper these functions are determined by comparing chains with length $\ell=19$ and $\ell=20$. $\beta(E)$ can be determined as $\beta(E)=\mathrm{ln}\{ \tau_{19}(E)/\tau_{20}(E) \}$, and $\alpha(E)$ is then given by $\alpha(E)=\tau_{20}(E)\mathrm{exp}\{ 20\beta(E) \}$. The $\alpha(E)$ and $\beta(E)$ for the system in Fig.~\ref{bindinggroup} a) are shown in Fig.~\ref{beta_alpha_pedagogic}. As the Hamiltonian for the chain is particle-hole symmetric with respect to $E=0$ for the chosen set of parameters, it is no surprise that $\beta(E)$ inherits this trait. $\beta(E)$ has its maximum at $E_0=0$ where the transport is maximally suppressed. We see that $\beta(E)$ is the same for the two values of $\Gamma$ and conclude that $\beta(E)$ is a pure backbone property and is insensitive to details in the contact coupling $\Gamma$. Simulations, not shown, confirm that changes in $t_{Au-bind}$ also have no effect on $\beta$. $\alpha(E)$, on the other hand, is determined by the contact couplings, and the peaks in $\alpha(E)$ correspond to the binding group transmission resonances observed in Fig.~\ref{bindinggroup} a). These peaks are also found in the local density of states (not shown) at the binding group orbitals. This clearly indicates that it is the binding group that determines $\alpha(E)$. Thus, it is clear from Eq.~(\ref{S2}) that the transmission resonances caused by the binding group affect the length-independent contribution to the thermopower.

One piece of important information that can be inferred from thermopower measurements is whether the transport is HOMO- or LUMO-dominated. In this context it is important to define what we mean by HOMO and LUMO states. Are we restricted to the states extending over the backbone of the chain or do we also include the states localized at the interface of the binding group and gold? The interface states often sit in the HOMO-LUMO gap of the periodic backbone and can, as we have seen, dominate transport in this gap. When the transport is dominated by a resonance below (above) $E_F$, the thermopower is positive (negative), and we say that the transport through the chain plus interface is HOMO- (LUMO-) dominated. This does not necessarily translate to HOMO- (LUMO-) dominated transport through the backbone of the chain itself, it can also be related to an increased density of states at the interface.\cite{Prodan2007} The charge transport along the backbone is governed by $\beta(E)$. For $E_F<E_0$ ($E_F>E_0$), the HOMO- (LUMO-) band of the backbone dominates transport, which corresponds to a thermopower increasing (decreasing) with length, see Eq.~(\ref{S2}). The type of charge carrier that dominates the transport along the backbone should therefore be inferred from the length dependence of the thermopower.  For long enough chains, $\ell>\frac{\alpha'(E_F)}{\alpha(E_F)\beta'(E_F)}$, $S(E)$ is dominated by the first term of Eq.~(\ref{S2}). In this case, it is the band structure of the chain that determines the nature of the transport, and contact and binding group states can be neglected. We believe that this conclusion is important as the length dependence of the thermopower has previously caused some confusion in the literature. It should be noted that if HOMO- and LUMO-states contribute equally to transport the thermopower saturates with length. In this case the length-independent contribution to the thermopower determines its sign also in the long chain limit. In Eqs.~(\ref{tau}) and (\ref{S2}) this corresponds to $\beta'(E_F)=0$.

\subsection{Explaining the experimental trends}

Up to this point, we have mainly been concerned with understanding how the different parameters of the model affect transport. We will now use this understanding to fit our model to experimental observations, i.e. the measurements of alkanedithiols\cite{Li2008,Malen2009} and the measurements of alkane chains coupled directly to gold.\cite{Widawsky2013} The latter system corresponds to choosing $E_{bind}=E_C$ in our model. We will see that a single set of parameters is able to achieve very good agreement with these measurements. The motivation for fitting to experiments is threefold: First, it will let us rationalize the trends observed in the experimental data. Second, reproducing several independent experimental observations tests the validity of the model. Third, the fitted model might serve as a tool to analyze future measurements.

We start by determining the parameters of the chain. We define the onsite energy of the carbon atoms to be $E_C=0$, and measure all other energies relative to this. Photoemission spectroscopy measurements of alkanethiol monolayers, with $\ell=18$, on Au have determined the HOMO-LUMO gap to $E_{H-L}=7.85$~eV.\cite{Qi2011} We assume that this corresponds to the HOMO-LUMO gap of the backbone of the chain. Weak signatures of what could be binding group and contact states contributed to the photoemission in the gap.\cite{Qi2011} In the long chain limit our model results in $E_{H-L}=2(t_a-t_b)$, where $E_{H-L}$ is defined as the distance between the highest eigenvalue of the HOMO-band and the lowest eigenvalue of the LUMO-band. As $\ell=18$ is quite close to the long chain limit $2(t_a-t_b)\lesssim7.85$~eV is reasonable. The ratio $t_b/t_a$ can then be determined from conductance measurements. A larger ratio corresponds to a larger conductance. From the measurements of alkanedithiols\cite{Li2008} and alkanes directly coupled to gold,\cite{Widawsky2013} we obtain $t_a=9$~eV and $t_b=5.25$~eV.

DFT-calculations suggest that the 6s- and 6p-orbitals of gold give an increased transmission around $2$~eV above $E_F$.\cite{Wu2013} Photoemission measurements indicate that also the 5d-orbitals should give a prominent contribution to conductance in this energy range.\cite{Sekiyama2010} To represent this we choose $E_{Au}=2$~eV. The thiol group, on the other hand, is known to give a transmission resonance below $E_F$,\cite{Li2008} and we choose the onsite energy of sulfur to $E_S=-1$~eV.

For simplicity, we treat the sulfur like carbon atoms in the alkanedithiols, with the only difference that the onsite energy differs. The coupling between the two sulfur orbitals on an atom is thus given by $t_{S-S}=t_b$, and the coupling between sulfur and carbon is given by $t_{S-C}=t_a$. 

Now, we are only left to determine the coupling between the end gold atoms and the contacts, $\Gamma$, as well as the gold-carbon and gold-sulfur couplings. The choice of these parameters should reproduce the conductance and thermopower measurements.\cite{Li2008,Malen2009,Widawsky2013} We find that $\Gamma=4$~eV, $t_{Au-S}=-2$~eV, and $t_{Au-C}=-4$~eV perform well in this regard. All parameters can be found in Table~\ref{parameters}.

\begin{table}[ht]
\caption{Parameters of the model in units of eV, determined from comparison with experimental measurements.}
\begin{center}
\begin{tabular}{|r|l|}
  \hline
  $E_C$ & 0 \\ \hline
  $E_{Au}$ & 2 \\ \hline
  $E_S$ & -1 \\ \hline
  $t_a$ & -9 \\ \hline
  $t_b$ & -5.25 \\ \hline
  $t_{Au-C}$ & -4 \\ \hline
  $t_{Au-S}$ & -2 \\ \hline
  $t_{S-S}$ & $t_b$ \\ \hline
  $t_{S-C}$ & $t_a$ \\ \hline
  $\Gamma$ & 4 \\
  \hline
\end{tabular}
\end{center}
\label{parameters}
\end{table}

\begin{figure}[ht]
\begin{center}
{\resizebox{!}{80mm}{\includegraphics{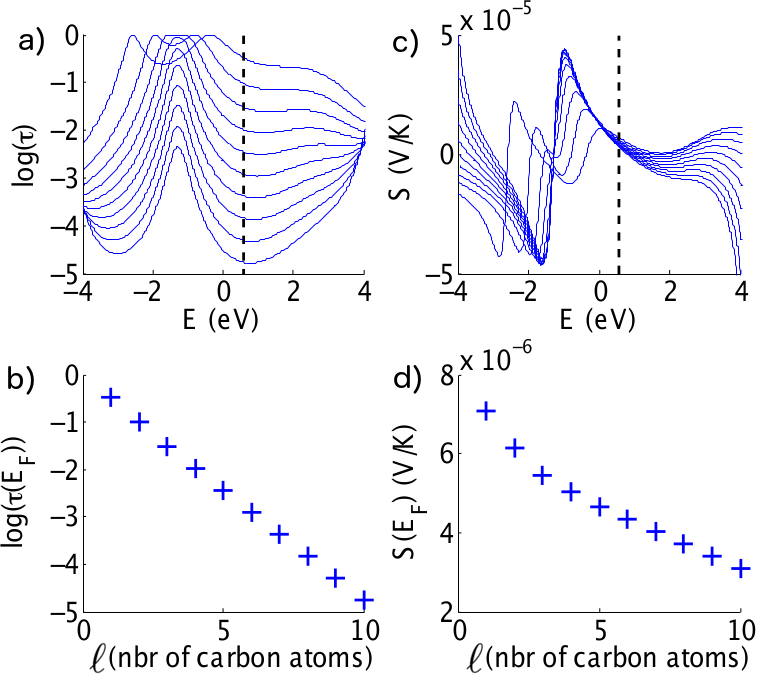}}}
\end{center}
\caption{a) $\mathrm{log}(\tau(E))$ for alkanedithiol chains with $\ell=1...10$ and parameters from Table~\ref{parameters}. The dashed line marks $E_F$ determined from comparison with experiments. b) $\mathrm{log}(\tau(E_F))$ as a function of chain length for the transmission of Fig.~a). c) $S(E)$ corresponding to the transmission in Fig. a), calculated using Eq.~(\ref{S}). d) $S(E_F)$ as a function of chain length.}
\label{exp_thiol} 
\end{figure}

The transmission functions for alkanedithiol with $\ell=1...10$ are shown in Fig.~\ref{exp_thiol}~a). The dashed line is an estimate of the Fermi level $E_F=0.57$~eV, relative to the onsite energy of the carbon atoms, obtained from comparison with experiments. The distance to the HOMO- and LUMO-band edge is then given by $E_F-E_H=4.51$~eV and $E_L-E_F=3.36$~eV, respectively. Not only does this estimate reproduce the conductance and thermopower measurements,\cite{Li2008,Malen2009} it is also in excellent agreement with the the Fermi level estimated from photoemission spectroscopy.\cite{Qi2011} The transmission resonance around $E=-1.3$~eV, seen in Fig.~\ref{exp_thiol}~a), originates from the thiol with onsite energy $E_S=-1$~eV. The slight shift towards more negative energies is a result from the hybridization with gold. A similar resonance was previously observed, see Fig.~7B) in Li et al..\cite{Li2008} The weak resonance roughly $2$~eV above $E_F$ originates from the gold atoms, in agreement with previous observations.\cite{Wu2013,Sekiyama2010} For short chains, the transmission is not exponentially suppressed at all energies. Instead, $\tau(E)\approx1$ at  the transmission resonances.

The exponential decrease in conductance at the Fermi energy is clear from Fig.~\ref{exp_thiol}~b), showing $\mathrm{log}(\tau(E_F))$. Neglecting the first three data points, as the transmission is not perfectly exponentially suppressed in the short chain limit, the data corresponds to an exponential suppression per carbon atom of $\beta_N=1.07$, in good agreement with experimental observations, reporting values in the range $\beta_N=0.83...1.07$.\cite{Xu2003,Lee2005,Li2008,Li2006}

We now change focus to the thermopower, which can be calculated from the transmission using Eq.~(\ref{S}) yielding the result shown in Fig.~\ref{exp_thiol} c). At the thiol resonance, the transmission is not exponentially suppressed for short chains. Consequently no clear trend is observed in the length dependence of the thermopower at these energies. At $E_F$, the thiol transmission resonance results in a negative value of $\alpha'(E_F)$ corresponding to a positive contribution to $S(E_F)$. As $E_F>0$ the transport is LUMO-dominated along the chain, which results in a thermopower decreasing with length. This trend is shown in Fig.~\ref{exp_thiol}~d), in good agreement with measurements.\cite{Malen2009} From the length dependence of the transmission it is hard to see if transport is HOMO- or LUMO-dominated along the chain, although we know that it is LUMO-dominated as $E_F>0$ resulting in $\beta'(E_F)<0$. LUMO-dominated transport is, however, easily inferred from the length dependence of the thermopower. The nonlinear deviation for the shortest chains results from the lack of exponential suppression of the transmission for these chains and $S(E_F)$ can thus not be written on the form of Eq.~(\ref{S2}) for chains with length $\ell=1...3$.

\begin{figure}[ht]
\begin{center}
{\resizebox{!}{80mm}{\includegraphics{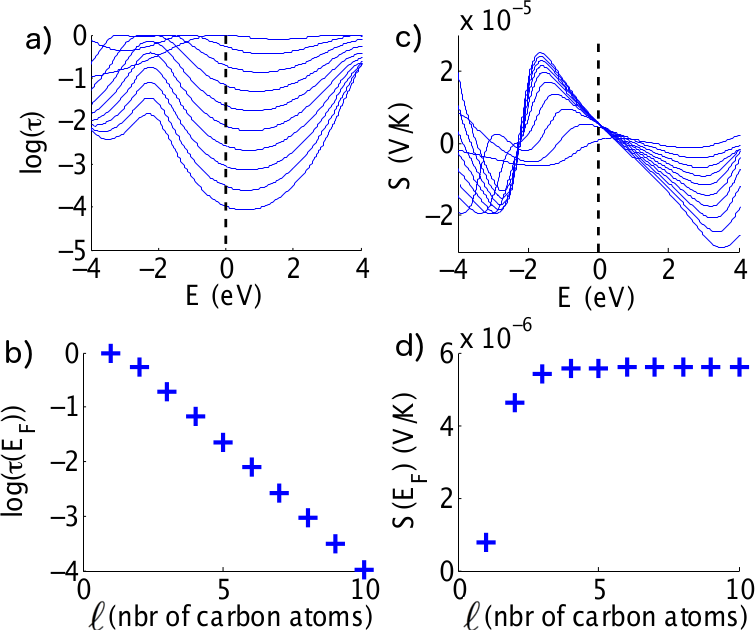}}}
\end{center}
\caption{a) $\mathrm{log}(\tau(E))$ for alkane chains directly coupled to gold with $\ell=1...10$ and parameters from Table~\ref{parameters}. The dashed line marks $E_F$ determined from comparison with experiments. b) $\mathrm{log}(\tau(E_F))$ as a function of chain length for the transmission of Fig.~a). c) $S(E)$ corresponding to the transmission in Fig. a), calculated using Eq.~(\ref{S}). d) $S(E_F)$ as a function of chain length.}
\label{exp_directAu} 
\end{figure}

The transmission for the alkane chains coupled directly to gold are shown in Fig.~\ref{exp_directAu}~a). Here the constant thermopower\cite{Widawsky2013} suggest that the Fermi level is situated in the middle of the HOMO-LUMO gap and $E_F=0$ is therefore assumed. It is not surprising that the position of $E_F$ relative to $E_C$ depends on the coupling to the contacts. The binding group determines the charge transfer onto the molecule thus affecting the band alignment of the system.\cite{Xue2001,Stadler2006,Peng2009,Balachandran2012}

A transmission resonance can be observed slightly below $-2$~eV. This originates from hybridization between eigenstates of the chain around $E=0$, and states of the gold. The strong coupling between the gold and the chain $t_{Au-C}=-4$~eV, supported by experimental observations,\cite{Widawsky2013} moves this resonance from $E=0$ to negative energies. Similar results was previously observed using DFT-simulations.\cite{Widawsky2013} In Fig. S11 b of the supplementary material of Widawsky et al.\cite{Widawsky2013} the transmission functions are shown for alkanes coupled directly to gold. A resonance similar to what was observed by us is found around $1$ eV below $E_F$. The gold resonance, appearing as a weak shoulder around $E=2$ in Fig.~\ref{exp_thiol} a) for  the alkanedithiols, is not seen in Fig.~\ref{exp_directAu}~a). Here the strong interaction between the gold and the chain shifts the gold resonance into the band of LUMO-states resulting from the periodic backbone of the chain.

\begin{figure}[h!]
\begin{center}
{\resizebox{!}{90mm}{\includegraphics{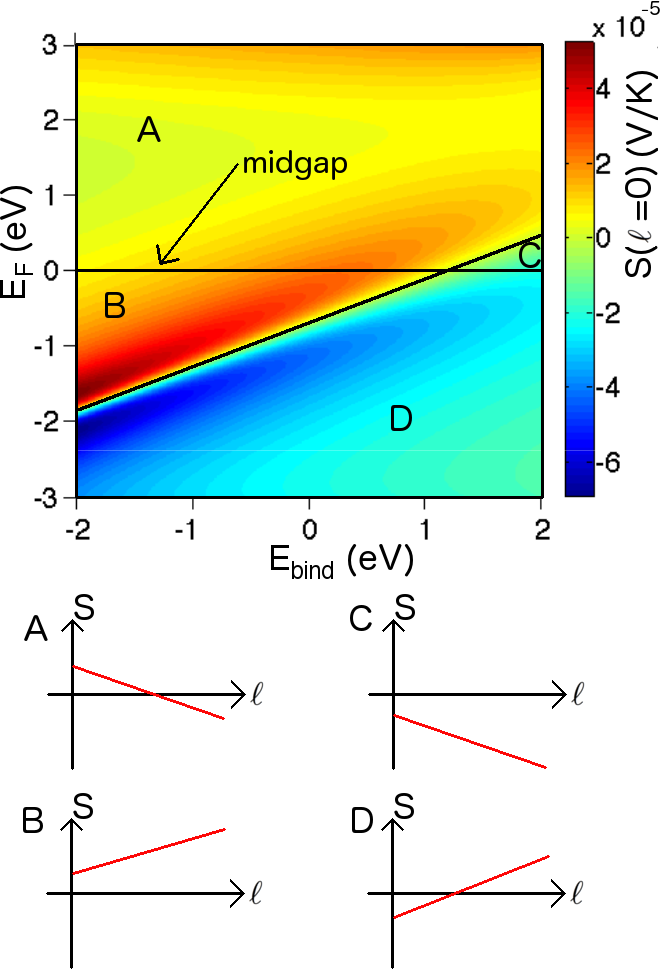}}}
\end{center}
\caption{$S(\ell=0)=-\frac{\pi^2k_B^2T}{3e}\frac{\alpha'(E)}{\alpha(E)}$ plotted as a function of the binding group energy $E_{bind}$ and the Fermi energy $E_F$, in the upper part of the figure. The black horizontal line marks the midgap $E=0$, $E_F<0$ ($E_F>0$) corresponds to HOMO- (LUMO-) dominated transport along the backbone of the chain, i.e. a thermopower increasing (decreasing) with length. The black diagonal line corresponds to $S(\ell=0)=0$. Above (below) this line $S(\ell=0)$ is positive (negative). The length dependence of the thermopower in the respective regions is sketched in the lower part of the figure. Parameters (except $E_{bind}$) are those of alkanedithiol taken from Table~\ref{parameters}.}
\label{Phase_diagram} 
\end{figure}

The exponential decrease in conductance is shown in Fig.~\ref{exp_directAu}~b), in very good agreement with experiments.\cite{Widawsky2013} The exponential suppression per carbon atom is slightly larger than for the alkanedithiols, $\beta_N=1.08$, as $E_F$ is positioned in the middle of the HOMO-LUMO gap. 

The thermopower corresponding to the transmission in Fig.~\ref{exp_directAu}~a) is shown in Fig.~\ref{exp_directAu}~c). It does not show a linear length dependence for short chains around $E_F$. This is more clearly seen in Fig.~\ref{exp_directAu}~d). We see that the thermopower saturates at a constant value as $E_F$ is in the middle of the HOMO-LUMO gap. The nonlinear length dependence for short chains can again be explained by that the thermopower can not be written on the form of Eq.~(\ref{S2}) in this regime. A thermopower saturating with length has been observed for benzene chains.\cite{Widawsky2013} The corresponding measurements for alkane chains also showed a slight nonlinear length dependence. We conclude that a thermopower saturating with length can be explained by a non-exponential length dependence of the conductance for short chains, in combination with a Fermi  level close to the middle of the HOMO-LUMO gap.

Having determined reasonable parameters for the model, it is of interest to see what thermopower trends are to be expected for different binding groups. For that purpose we plot $S(\ell=0)=-\pi^2k_B^2T(\alpha'(E)/\alpha(E))$ as a function binding group energy $E_{bind}$ and the Fermi energy $E_F$, in Fig.~\ref{Phase_diagram}. All parameters (except $E_{bind}$) are the ones for alkanedithiol from Table~\ref{parameters}. The black horizontal line marks the mid gap $E=0$, $E_F<0$ ($E_F>0$) corresponds to HOMO- (LUMO-) dominated transport, that is a thermopower increasing (decreasing) with length. The diagonal line separates the regions where $S(\ell=0)>0$ (above) and $S(\ell=0)<0$ (below). These two lines result in four regions with qualitatively different trends in the thermopower, which is either positive or negative in the short chain limit, and either decreases or increases with length. The length dependence of the thermopower in the respective regions is sketched in the lower part of the figure.

As previously discussed $E_{bind}$ affects the position of the binding group induced transmission resonance, which is important for $S(\ell=0)$. In Fig.~\ref{Phase_diagram}, the diagonal line corresponds to $\alpha'(E)=0$, which happens right at the binding group induced transmission resonance. Due to the coupling to gold with onsite energy $E_{Au}=2$~eV the transmission resonance is shifted to lower energies. This causes the asymmetry in the figure where the areas with $S(\ell=0)>0$ are larger. For systems with larger $t_{Au-bind}$ such as alkanes directly coupled to gold the shift would be larger and the asymmetry is more pronounced.

Fig.~\ref{Phase_diagram} could be thought of as a rough roadmap to what thermopower trends are to be expected for different structures. Alkanedithiols fall into area A according to Malen et al.,\cite{Malen2009} while alkanes coupled directly to gold sit at the border between A and B.\cite{Widawsky2013} Since the thermopower can always be written on the form of Eq.~(\ref{S2}), as long as the transport is exponentially suppressed, similar roadmaps are to be expected for chains with different backbones such as phenylenes. These chains fall into region B when coupled to gold via thiols or amines.\cite{Malen2009} The measurements of benzene coupled via cyano groups\cite{Baheti2008} show that this binding group results in a negative contact thermopower. The length dependence has, however, not been measured so one cannot say whether the results would end up in region C or D.

\begin{figure}[h!]
\begin{center}
{\resizebox{!}{100mm}{\includegraphics{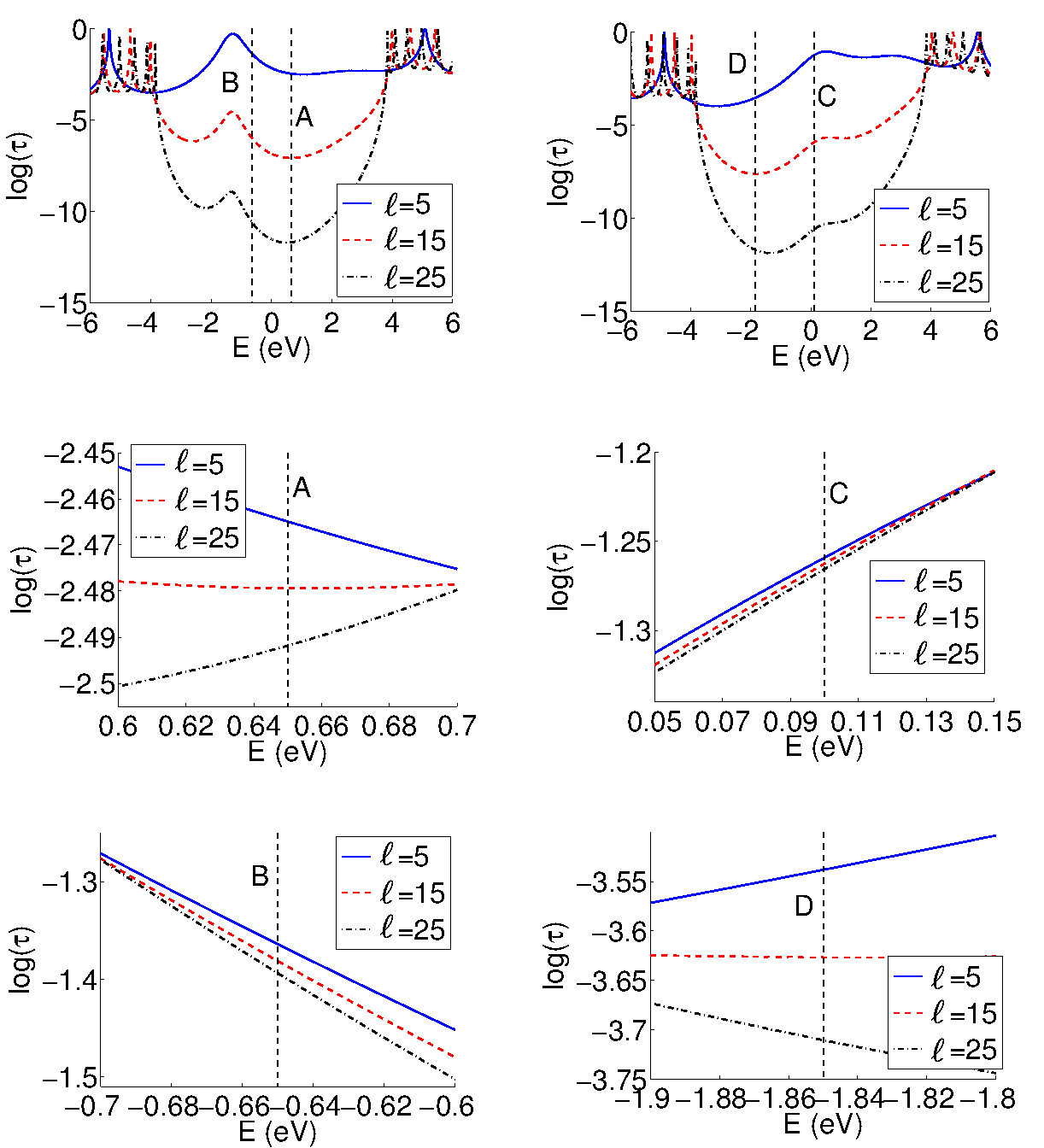}}}
\end{center}
\caption{Transmission functions corresponding to different $E_{bind}$ in Fig.~\ref{Phase_diagram}. The left panel shows the transmission functions for $E_{bind}=-1$~eV where $S(\ell=0)>0$, while the right panel shows the transmission functions for $E_{bind}=2$~eV where $S(\ell=0)<0$. The vertical dashed lines mark the positions of $E_F$ corresponding to the areas A...D in Fig.~\ref{Phase_diagram}. The upper two figures show the transmission functions for chain lengths $\ell=5, 15$ and $25$. The lower four figures show the results around the vertical dashed lines. Here the transmission functions for $\ell=15$ and $25$ are shifted vertically (by adding a constant to $\mathrm{log}(\tau)$) for easier comparison.}
\label{E_bind} 
\end{figure}

We have seen that the binding group results in a transmission resonance. To improve the chemical intuition for $E_{bind}$ we show in Fig.~\ref{E_bind} how $E_{bind}$ affects the transmission functions at different chain lengths $\ell=5, 15$ and $25$. The left panel of the figure shows the transmission functions for $E_{bind}=-1$~eV where $S(\ell=0)>0$, while the right panel shows the transmission functions for $E_{bind}=2$~eV where $S(\ell=0)<0$. The vertical dashed lines mark the positions of $E_F$ corresponding to the areas A...D in Fig.~\ref{Phase_diagram}. The upper two figures show the transmission functions for chain lengths $\ell=5, 15$ and $25$. The lower four figures show the results around the vertical dashed lines. Here the transmission functions for $\ell=15$ and $25$ have been shifted vertically (by adding a constant to $\mathrm{log(\tau)}$) for easier comparison. We see that around the vertical line A, the slope of the transmission is negative for short chains, due to the transmission resonance from the binding group. For increasing chain lengths this transmission resonance is exponentially suppressed and transport is instead dominated by the LUMO-states of the periodic backbone. This results in a thermopower which is positive for short chains and negative for longer chains. At the vertical line B, transport is instead dominated by the HOMO-states of the periodic backbone for longer chains. Thus, the slope of the transmission in negative for all chain lengths, corresponding to a positive thermopower. For $E_{bind}>0$, i.e. the right panel of the figure, the reasoning is very similar. Here it should, however, be noted that the position of the transmission resonance is quite different from $E_{bind}=2$~eV. The reason for this is that here $E_{bind}$ coincides with $E_{Au}=2$~eV. These states hybridize and form a bonding and an antibonding state. The bonding states results in a transmission resonance slightly above $E=0$, while the antibonding state can be seen as a weak shoulder in the transmission for $\ell=5$.

The results from Malen et al.\cite{Malen2009} and Guo et al.\cite{Guo2013} are qualitatively different, albeit there is no obvious difference between the systems. However, contact geometries necessarily vary from junction to junction. This can, as we have seen, affect the length independent contribution to the thermopower, as well as its length dependence via the relative position of the Fermi energy to the molecular orbitals. It is very interesting that such effects can result in qualitatively different trends, moving the alkanedithiols from area A to B or even D in Fig.~\ref{Phase_diagram}. This demonstrates the importance of properly analyzing thermopower measurements, as they might hold detailed information about the junction geometry.

\section{Methods}

Although the model allowed for good quantitative agreement with experiments, the main objective was to explain experimentally observed trends. For that purpose, we adopted an approach where the results could be easily interpreted. Despite its simplicity, the model included all features required to explain the observed trends.

We will end by a short discussion of effects neglected in the model. Such an effect is next-nearest neighbor coupling along the chain. Without next-nearest neighbor coupling the HOMO- and LUMO-bands from the periodic backbone of the chain have equal width and $\beta(E)$ is symmetric around the middle of the HOMO-LUMO gap. Introducing next-nearest neighbor coupling breaks this symmetry so that $E_0$, defined as $\mathrm{arg}\{\mathrm{max}(\beta(E))\}$, is shifted away from $E_0=0$. However, it is not obvious how to determine the degree of asymmetry from existing experimental data, and we have therefore chosen not to include this effect. The asymmetry could be probed experimentally by measuring the the length dependence of the transmission for different gate voltages. This would allow for determination of $\beta(E)$ over an energy range, and not only at the Fermi energy of the un-gated structure.

Another mechanism not included in our approach is sequential hopping along the chain. For longer chains this is the dominant transport mechanism, resulting in that the conductance is no longer exponentially suppressed.\cite{Segal2005} However, experiments indicate exponential suppression at least for alkane chains up to $10$ carbon atoms.\cite{Li2008}

\section{Conclusion}

We conclude that the thermopower decreasing with chain length,\cite{Malen2009} as well as the constant thermopower,\cite{Widawsky2013} can be understood in the following way: the length dependence of the thermopower depends on the position of the Fermi level. A decreasing thermopower corresponds to the LUMO-band, from of the states extending over the backbone of the chain, dominating transport. A constant thermopower, on the other hand, corresponds to equal contributions from electrons and holes to the transport along the backbone. The short chain thermopower is, on the other hand, determined by the transmission resonances due to the binding group and contact states.

We can see that with the careful exercise of judgement that Hammett mentioned, we can recover the understanding that increasing the length of an alkane chain can indeed produce similar changes in the thermopower. It is also clear, however, that a naive application of this assumption will not hold as the combination of the length of the chain, the binding group and the position of the Fermi energy work together to yield the trend we observe. We are most certainly in the regime of \textit{``wisdom that comes only with experience''} and should bear this in mind as we search for structure-function relationships in physical observables outside the traditional domain of chemistry. 

\begin{acknowledgments}
The research leading to these results has received funding from the European Research Council under the European Union's Seventh Framework Program (FP7/2007-2013)/ERC Grant Agreement No. 258806. 
\end{acknowledgments}

%\bibliography{bibfileJCP}

%

\end{document}